# Quantifying the Social Costs of Power Outages and Restoration Disparities Across Four U.S. Hurricanes

Xiangpeng Li[1*], Junwei Ma[1], Bo Li[1], Ali Mostafavi[2]

[1] Ph.D. student. Urban Resilience.AI Lab, Zachry Department of Civil and Environmental Engineering, Texas A&M University, College Station, Texas, United States.

[2] Professor. Urban Resilience.AI Lab, Zachry Department of Civil and Environmental Engineering, Texas A&M University, College Station, Texas, United States.

[*] Corresponding author: Xiangpeng Li, E-mail: xplli@tamu.edu.

**Abstract**

The multifaceted nature of disaster impact reveals that areas with concentrated populations may contribute more to overall economic burden, while sparsely populated but highly impacted regions suffer disproportionately at the individual level. This study presents a framework that quantifies the societal costs of power outages by converting customer-weighted outage exposure into monetized social losses. This framework integrates welfare metrics with three recovery measures—average outage-days per customer, restore duration, and relative restoration rate—computed from sequential EAGLE-I observations and linked to Zip Code Tabulation Area-level demographics. Through a standardized pipeline applied to four U.S. hurricanes—Beryl (2024, Texas), Helene (2024, Florida), Milton (2024, Florida), Ida (2021, Louisiana), we produce the first cross-event, fine-scale accounting of societal outage costs and their underlying drivers. The analysis reveals substantial heterogeneity in impacts: Ida generated $1.50 billion in total deprivation costs ($1,757 per capital) Milton, $1.26 billion ($387 per capita); Beryl, $629 million, ($674 per capita); and Helene, $411 million ($285 per capita). Across all events, the results document consistently regressive burden distributions, with deprivation-cost shares declining systematically with income. Mechanistic analysis demonstrates that deprivation costs increase with restoration duration and decrease with relative restoration rates. Explainable modeling identifies restore duration as the dominant predictor of costs, while unsupervised clustering reveals distinct recovery typologies invisible to conventional reliability metrics. By shifting from simple outage counts and subjective assessments to welfare-based, exposure-weighted measurement, this study delivers three key contributions: (I) a transferable framework for quantifying outage impacts and equity, (ii) comparable cross-event evidence linking restoration dynamics to social losses, and (iii) actionable spatial analyses and community typologies that enable equity-informed restoration prioritization and resilience investment decisions.

**Introduction**

Power systems represent critical infrastructure whose failure can trigger cascading societal and economic disruptions[1–3], yet current approaches to evaluating grid performance during extreme weather events remain fundamentally inadequate[4]. Traditional utility metrics, such as System Average Interruption Duration Index (SAIDI) and Customer Average Interruption Duration Index (CAIDI), provide aggregate system-level indicators[5,6] but fail to capture the heterogeneous impacts experienced by different communities during major outages. These conventional measures treat all customers equally, obscuring the reality that power interruptions impose vastly different burdens depending on household income, social vulnerability, and local infrastructure conditions. The increasing frequency and intensity of extreme weather events driven by climate change have intensified scrutiny of electric grid resilience and recovery performance[7,8]. Recent hurricanes have demonstrated that power outage impacts extend far beyond simple restoration timelines, generating profound economic losses and exacerbating existing social inequalities[9,10]. However, the methodological frameworks for quantifying these broader societal costs remain underdeveloped, limiting both academic understanding and policy responses to infrastructure vulnerability.

Current limitations in outage impact assessment stem from several critical methodological deficiencies. Most significantly, existing studies fail to implement deprivation cost frameworks that can translate outage duration into meaningful economic welfare losses experienced by affected populations[11–13]. While researchers have developed theoretical foundations for quantifying infrastructure service deprivation[14,15], practical applications to power outage analysis remain limited, preventing a comprehensive assessment of societal burden. Additionally, the reliance on aggregate utility performance metrics provides insufficient granularity to identify communities experiencing disproportionate impacts[14], and the absence of standardized methodologies for customer-weighted duration calculations constrains accurate measurement of cumulative exposure across heterogeneous populations. Also, the existing literature on power system resilience exhibits a critical disconnect between engineering-focused performance metrics and the actual societal costs of outages[16,17]. Traditional assessments overwhelmingly rely on aggregate indices (e.g., SAIDI, CAIDI) that obscure the spatially heterogeneous and inequitable distribution of impacts by treating all customers uniformly[18,19]. A significant deficiency is the failure to systematically translate physical service interruptions into quantifiable economic welfare losses experienced by diverse households[20]. While theoretical frameworks for deprivation costs exist, their practical application to large-scale, cross-event analysis remains nascent, preventing a comprehensive assessment of the true societal burden of grid failure and limiting the ability of policymakers and utilities to address social equity in restoration strategies. This study departs from these limitations by operationalizing a comprehensive deprivation cost framework that integrates high-resolution temporal outage data with fine-scale socioeconomic data. We provide the first systematic quantification and comparison of equity outcomes across four major hurricane events, utilizing advanced analytical techniques (Shapley Additive Explanations and spatial clustering) to uncover the underlying drivers and hidden patterns of disproportionate societal costs that conventional approaches completely overlook.

Accordingly, this study tackles three linked questions central to outage equity and decision-making: (1) To what extent do societal costs of power outages fall disproportionately on lower-income communities? (2) What drives variation in deprivation costs—restoration dynamics (duration and rate) or baseline sociodemographic? (3) Can clustering techniques identify recovery patterns that conventional reliability indices overlook but that are critical for resource prioritization? To answer these, we compute Zip Code Tabulation Area (ZCTA)-level, customer-weighted outage exposure from sequential observations and translate it through an empirically estimated deprivation-cost function, enabling cross-event comparability (Beryl, Helene, Milton, Ida) and equity-focused inference. Framing the inquiry this way separates who is most burdened (income-stratified impacts), why (mechanisms tied to restoration speed and duration), and recovery actions to undertake (clusters that pinpoint high-burden, slow-recovery communities), turning descriptive outage statistics into decision-ready evidence for regulators and utilities.

By applying this framework to four major hurricane events—Beryl (Texas), Helene (Florida), Milton (Florida), and Ida (Louisiana)—we provide the first systematic comparison of outage equity outcomes across different regions and storm characteristics. Through spatial clustering analysis of deprivation costs, restoration performance, and socioeconomic indicators, we identify distinct community typologies that reveal how infrastructure vulnerability intersects with social vulnerability to produce systematically different disaster outcomes. The research addresses three critical questions: to what extent do societal costs of power outages disproportionately affect low-income areas; how do outage characteristics and socio-demographic features contribute to deprivation costs; and to what extent does clustering analysis reveal hidden heterogeneity in outage burden that conventional metrics overlook.

This study advances outage-impact assessment by pricing deprivation rather than merely counting interruptions: it converts customer-weighted outage-days into household welfare losses via an empirically derived deprivation-cost function and pairs that metric with three recovery measures—restore duration, relative restoration rate, and average outage-days per customer—to capture both the pace and the fairness of restoration. Applying a single, reproducible pipeline to four hurricanes (Beryl–TX 2024; Helene–FL 2024; Milton–FL 2024; Ida–LA 2021) by linking high-frequency Environment for Analysis of Geo-Located Energy Information (EAGLE-I™) traces with ZCTA-level demographics yields the first cross-event, fine-scale accounting of societal outage costs that SAIDI/CAIDI cannot see; for example, Ida's losses approach \$1.5 B, while per-capita burdens are systematically higher in lower-income areas. Explainable modeling (SHAP) and K-means clustering then translate these measurements into operational insights—identifying the dominant drivers (especially restoration duration and comparative restoration speed) and revealing distinct recovery typologies that conventional metrics miss. By moving the discussion from reliability indices to welfare-based equity impacts, the framework gives regulators and utilities a transparent, decision-ready basis for equity-informed restoration prioritization and resilience investment, targeting reductions where delays are most socially costly.

Integrating deprivation cost functions with observational outage metrics transforms how we measure power disruption impacts, moving from proxy indicators to exposure-weighted, welfare-based assessment. Sequential EAGLE-I™ traces provide customer-weighted outage-days, restore duration, and relative restoration rates. When processed through our empirically derived valuation function, these metrics yield comparable dollar values of social loss—an additive measure that enables direct prioritization and cross-event benchmarking. This approach addresses fundamental limitations of traditional metrics. While extent and duration indices like SAIDI and CAIDI treat all customers equally, our framework captures both the duration of service loss and its differential social value, revealing inequities hidden in system-wide averages. Moreover, by leveraging complete, high-frequency observational data rather than sparse, retrospective surveys, we generate fine-scale, reproducible estimates that remain stable across storms and regions while avoiding recall and sampling biases. The framework provides utilities and regulators with a decision-ready tool that directly links restoration dynamics to monetized welfare losses. This enables the targeting of resources to communities where accelerated restoration or improved restoration rates prevent the largest social costs, rather than relying on simple customer counts or anecdotal evidence. By quantifying the welfare implications of each additional outage day across different communities, decision-makers can optimize both efficiency and equity in their response strategies.

**Study area and data**
This study examines power outage impacts across four major hurricane events that occurred between 2021 and 2024, encompassing diverse geographic, climatic, and socioeconomic contexts across the Gulf Coast and southeastern United States. The selected events—Hurricane Beryl (2024) in Harris County, Texas; Hurricane Helene (2024) in Florida; Hurricane Milton (2024) in Florida; and Hurricane Ida (2021) in Louisiana—represent varying storm intensities, landfall locations, and regional characteristics that enable comprehensive analysis of outage burden heterogeneity.

*Hurricane Beryl (2024, Harris County, Texas)*
Hurricane Beryl made landfall in Texas on July 8, 2024, causing widespread power outages across the Houston metropolitan area[21]. Harris County, encompassing Houston and surrounding communities, represents one of the most populous and economically diverse counties in the United States, with more than 2.7 million residents[22].

*Hurricane Helene (2024, Florida)*
Hurricane Helene struck Florida's Big Bend region on September 26, 2024, as a Category 4 storm, generating extensive power outages across the state[23]. Florida's geographic diversity, spanning coastal plains to inland agricultural areas, provides varied exposure conditions for analyzing outage impacts.

*Hurricane Milton (2024, Florida)*
Hurricane Milton made landfall on Florida's west coast on October 9, 2024, as a Category 3 storm[24], affecting many of the same communities impacted by Hurricane Helene just two weeks earlier. This temporal proximity provides a unique opportunity to examine cumulative outage effects and restoration system resilience under repeated stress.

*Hurricane Ida (2021, Louisiana)*
Hurricane Ida struck Louisiana on August 29, 2021, as a Category 4 storm, causing catastrophic power outages across southeastern Louisiana[25].

*Data Sources and Characteristics*
Power outage data for all events (Table 1) were obtained from the Environment for Analysis of Geo-Located Energy Information (EAGLE-I™) platform (U.S. Department of Energy). This platform, developed by Oak Ridge National Laboratory, serves as a geographic information system and data visualization tool, providing high-resolution customer outage data. The datasets include timestamps at 15-minute to hourly intervals, enabling fine-scale temporal analysis of outage development and restoration patterns. Each record contains the number of customers without power and total customers served, allowing calculation of outage percentages and absolute impact measures. Socioeconomic and demographic data were obtained from the U.S. Census Bureau's American Community Survey (ACS) 5-year estimates (2018-2022)[26], providing ZCTA-level information on median household income, population demographics, housing characteristics, and economic indicators. These data enable analysis of how pre-existing community conditions influence outage vulnerability and recovery outcomes. Geographic boundary data for ZCTAs were obtained from the U.S. Census Bureau's TIGER/Line shapefiles[27], enabling spatial analysis and mapping of outage impacts. The ZCTA geography provides an appropriate spatial scale for utility service analysis while maintaining sufficient resolution to capture neighborhood-level variation in impacts and outcomes.

Table 1. Data sources

| Data | Location | Source |
|---|---|---|
| Beryl power outage data | Harris County, Texas | EAGLE-I™ |
| Helene power outage data | Florida | EAGLE-I™ |
| Milton power outage data | Florida | EAGLE-I™ |
| Ida power outage data | Louisiana | EAGLE-I™ |
| Median Income | All | Census |
| ZCTA map | All | Census |

Figure 1 shows a comprehensive methodological framework of this study. The research integrates multiple data sources, including high-resolution power outage data from four hurricane events (Beryl, Helene, Milton, and Ida), socio-demographic information capturing median income characteristics, Zip code-level geometric boundaries, and an empirically derived deprivation cost function that translates outage duration into monetary losses based on household willingness to pay for restoration. The data processing stage transforms raw outage observations into meaningful metrics through affected area identification using threshold-based filtering, calculation of customer-weighted average outage duration, restoration duration measurement, and relative restoration rate computation that normalizes recovery speed against outage accumulation rates. The analytical framework employs curve fitting techniques to model relationships between income and deprivation costs, SHAP analysis to quantify feature contributions to societal cost, and K-means clustering to identify distinct community typologies based on outage characteristics, deprivation cost, and socioeconomic indicators. The result analysis synthesizes findings across three dimensions: examining relationships between income and deprivation burden, quantifying how outage features and demographic characteristics contribute to societal costs, and revealing hidden heterogeneity in community experiences that conventional utility performance metrics fail to capture, ultimately providing a comprehensive assessment of equity outcomes in power system resilience.

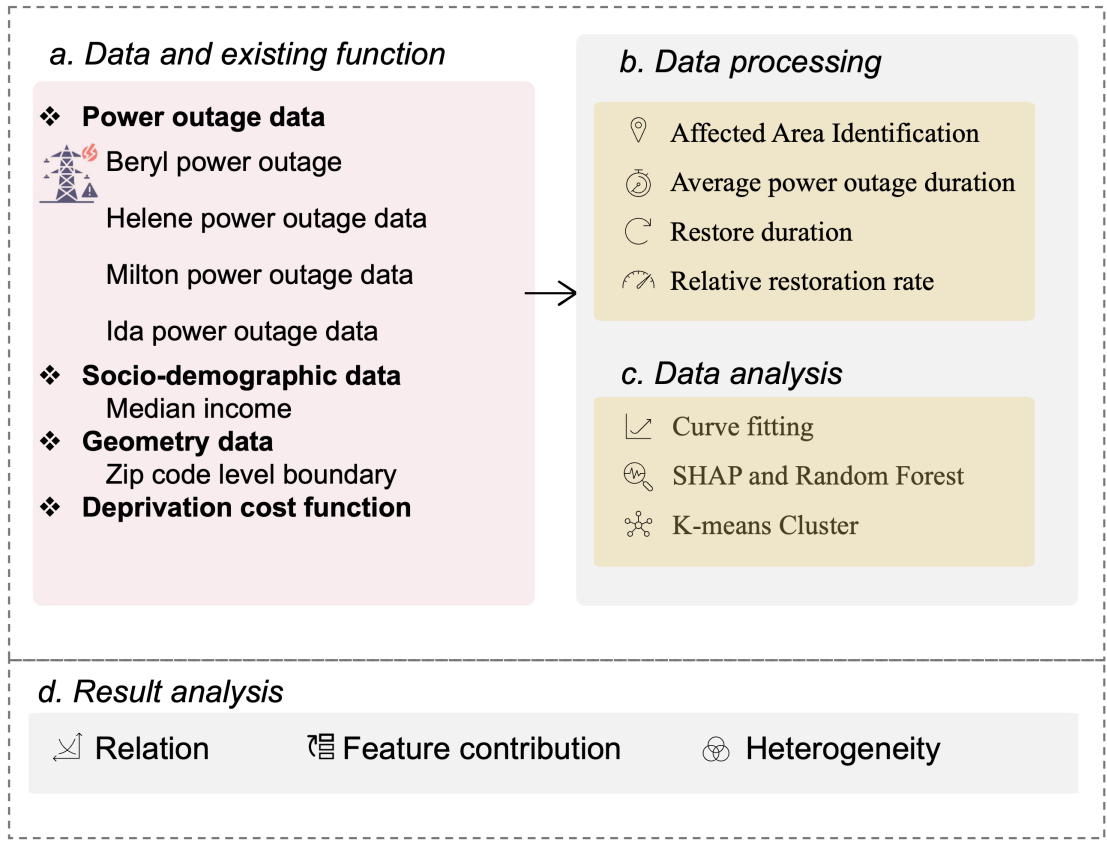

**Figure 1. Overview of the study**

**Method**

*Affected Area Identification*

Because Florida and Louisiana are state-level events, the hurricanes did not impact every area uniformly. To focus on the locations affected by power outages, we applied a data-driven filtering approach to identify the specific Zip Code Tabulation Areas (ZCTAs) impacted during each hurricane event (Helene, Milton, and Ida). For all events, we compared each ZCTA's median outage percentage during the defined impact window to its baseline value and classified the ZCTA as affected if the impact median exceeded the threshold. For Helene and Milton, the baseline period was defined as September 19–25, 2024, representing typical pre-storm outage conditions. For Hurricane Ida, due to the absence of pre-event data, the baseline period was defined using the after-power outage window from September 18–24, 2021. To account for the elevated outage levels inherent in this recovery-based baseline, a more conservative threshold was applied: ZCTAs were classified as affected only if the median outage during the impact window exceeded the baseline median. For Ida, we set it as 80% of the baseline.

*Average power outage duration (days) per customer*
The approach processes sequential outage data to determine how long each customer was without power. Customer-hours are computed by multiplying the number of affected customers by the estimated duration for each observation period. The calculation proceeds through two aggregation stages. First, customer hours are summed daily within each Zip code while tracking the maximum number of customers affected per day. Daily totals are then converted to outage-days by dividing by 24 hours. Second, outage days are summed across all days of the event for each Zip code, maintaining the overall maximum customer count throughout the event period. The final metric represents average outage days per customer, calculated by dividing total outage days by the maximum number of customers affected in each Zip code. This measure captures the mean duration of service interruption experienced by customers during the event, accounting for both the number of people affected and how long they remained without power.

*Deprivation cost function*
In the deprivation cost function (DCF)[28–32]: infrastructure failures generate deprivation—the acute absence of essential services—which consequently produces human suffering and economic consequences far exceeding direct monetary losses. The DCF framework fundamentally transforms infrastructure resilience investment evaluation by quantifying welfare losses through economic valuation, thereby capturing the full spectrum of societal impacts. Equation 1 shows the estimation of the deprivation cost during a power outage after a hurricane:

$$DC = 35.95 \cdot t^2 + 107.84 \cdot t + 71.89 \qquad \text{Eq.1}$$

where *DC* means the estimated monetary value of deprivation cost (in dollars), where *t* represents the average power outage duration (days) per customer. The total deprivation cost is the average *DC* times total affected customers.

*Restore duration*
The restore duration metric represents the length of time, in days, required for a Zip Code Tabulation Area to recover from a power outage following a major disruption. Specifically, it is defined as the interval between the onset of restoration—marked by the first day when customer restoration activity is observed—and the point at which the resilience curve reaches or exceeds zero. The resilience curve is constructed by tracking the cumulative number of customers restored against the cumulative number of customers affected. Restore duration therefore captures the active recovery phase, reflecting how long it takes for restoration efforts to meaningfully offset accumulated outages. This temporal measure provides insight into the pace and persistence of recovery efforts across different regions, serving as a key indicator of restoration performance following extreme weather events.

*Relative restoration rate*

To evaluate the efficiency of post-outage recovery across ZCTAs, we compute relative restoration rate. This metric captures the proportionate speed at which service is restored relative to the rate at which outages accumulate. Specifically, for each ZCTA, the outage rate is defined as the peak cumulative number of customers affected divided by the duration (in days) over which the outage developed. The restoration rate in equation 2 is similarly defined as the total number of customers restored divided by the duration (in days) of the recovery phase. The relative restoration rate is then calculated as the ratio of restoration rate to outage rate:

$$Relative\ Restoration\ Rate = \frac{Restoration\ Rate}{Outage\ Rate} = \frac{\frac{Total\ Customer\ Restored}{Restore\ Duration}}{\frac{Max\ Customer\ Outage}{Outage\ Duration}} \qquad \text{Eq.2}$$

A value greater than 1 indicates a recovery that was faster than the outage accumulation, while a value less than 1 suggests a lagging restoration process. This normalized metric enables cross-region comparisons of restoration performance regardless of outage magnitude.

*Curve fitting*

To investigate income-related disparities in power outage impacts, we conducted a cross-event comparison of deprivation cost relative to household income. Deprivation cost share was defined as the ratio of estimated monetary loss due to power outages to median household income for each geographic unit. Data from four major events—Hurricane Beryl (Texas), Helene (Florida), Milton (Florida), and Ida (Louisiana)—were collected and cleaned to retain only valid numeric entries with positive values. For each event, five candidate functional forms—linear, inverse, power, logarithmic-linear (loglin), and exponential decay—were fit using non-linear least squares regression. The best-fitting model for each event was selected based on the lowest Akaike Information Criterion (AIC)[33], balancing model complexity with explanatory power. In addition, the study also performed curve fitting analyses between deprivation cost and relative restoration rate, as well as between deprivation cost and restoration duration.

*K-means Cluster*

This study utilizes K-means cluster[34], an unsupervised machine learning algorithm, to identify groups of ZCTAs that exhibit similar characteristics in terms of outage exposure, socio-economic vulnerability, and restoration performance. The algorithm partitions the dataset into *k* clusters by minimizing the within-cluster sum of squared distances. Formally, given a set of observations $X = \{x_1, x_2, x_3, \ldots, x_n\}$, the K-means algorithm aims to find cluster centroids $\{\mu_1, \mu_2, \mu_3, \ldots, \mu_k\}$ that minimize the objective function:

$$argmin\ C \sum_{i=1}^{k} \sum_{x \in C_i} \|x - \mu_i\|^2 \qquad \text{Eq.3}$$

where $C_i$ denotes the set of points assigned to cluster *i*, and $\mu_i$ is the centroid of that cluster. Each data point is assigned to the nearest centroid, and centroids are iteratively updated until convergence. In this study, features including deprivation cost, average outage duration, median household income, and restoration metrics were standardized and used as input variables. The resulting clusters reveal underlying patterns and enable the classification of ZCTAs into distinct typologies of outage experience and recovery response.

*Random Forest and SHAP*

To analyze the drivers of deprivation costs, we employed a Random Forest model, a widely used ensemble learning method that builds multiple decision trees and aggregates their outputs to improve predictive accuracy and robustness. Random Forests are particularly well-suited for handling nonlinear relationships and interactions among predictors, making them appropriate for complex resilience data[35]. However, like many machine learning models, their interpretability is limited. To address this, we applied SHAP[36], a model-agnostic framework grounded in cooperative game theory, which decomposes each prediction into additive contributions from the input features. This combination allowed us not only to generate accurate predictions of deprivation costs but also to quantify the relative importance of income, restoration duration, and relative restoration rate, as well as to trace their specific contributions for individual regions.

**Results**

The event average deprivation cost and total deprivation cost is shown in Table 1. Hurricane Ida in Louisiana stands out as the most economically devastating event, responsible for nearly $1.5 billion in total deprivation costs. Florida experienced two major events—Helene and Milton—with markedly different cost profiles. While approximately $411 million in total costs was attributable to Helene, Milton's impact was more than three times greater at $1.26 billion, despite both affecting the same state. The average deprivation cost per affected individual reveals even more nuanced patterns. Louisiana's Hurricane Ida imposed the highest individual burden at $1,757 per person, indicating severe social cost per customer. Conversely, Hurricane Helene in Florida shows the lowest average deprivation cost at $285 per person, despite generating substantial total costs. The relationship between total and average costs provides insights into regional total social loss and average social loss. Harris County's experience with Beryl shows moderate total costs ($629 million) but relatively high per-person impacts ($674). Florida's ability to keep per-person costs relatively low for both storms, despite high total costs, indicate widespread affected customer but short total affected power outage duration.

**Table 2. Deprivation cost in different events**

| Event | Location | Total deprivation costs (USD) | Average deprivation costs (USD) |
|---|---|---|---|
| **Beryl** | Harris County | 629,261,779.00 | 674.09 |
| **Helene** | Florida | 410,932,627.05 | 285.42 |
| **Milton** | Florida | 1,259,261,698.13 | 387.01 |
| **Ida** | Louisiana | 1,496,970,026.06 | 1757.08 |

Figure 2 through Figure 5 present the spatial patterns of both average and total deprivation cost (DC) across Zip Codes in the four events, where left panel displays the *average deprivation cost per person*, representing the mean individual-level economic loss associated with power outage duration. The right panel illustrates the *total deprivation cost* per ZIP code, aggregating the individual costs to reflect the community-level burden.

In Figure 2, Northern and northeastern areas—particularly those along the urban–rural fringe—exhibit the highest average deprivation costs, often exceeding $1,600 per person. In contrast, central and western urban areas show lower average DCs. In terms of total deprivation cost, the northeastern and eastern suburban zones stand out with cumulative losses surpassing $30 million in some areas. To note, the central areas show high average deprivation cost but low total deprivation cost.

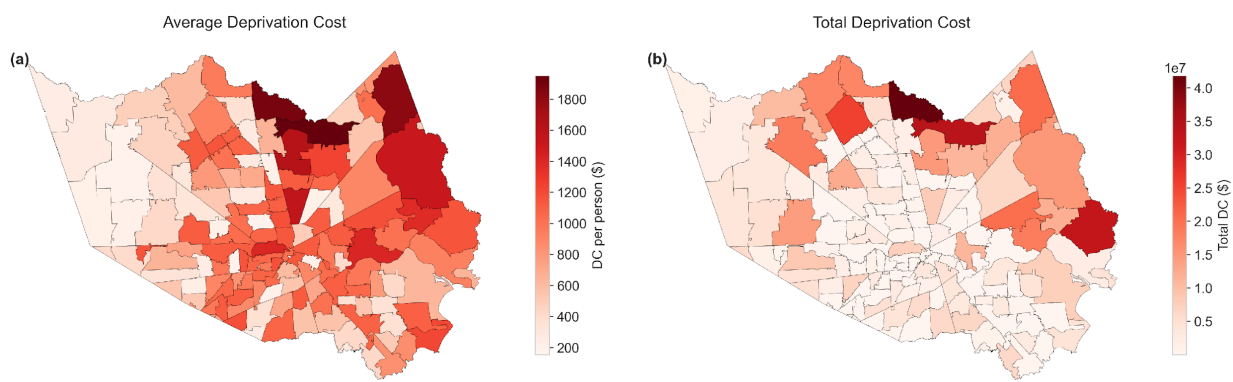

**Figure 2. Spatial distribution of power outage deprivation costs following Hurricane Beryl in Harris County, Texas**. (a) Average deprivation cost per person, showing individual-level economic burden with the highest costs (>$1,600) in northern and northeastern suburban areas. (b) Total deprivation cost per Zip Code, indicating community-level aggregate losses with peak values (>$30 million) in northeastern and eastern zones.

Figure 3 illustrates the spatial distribution of deprivation costs across Florida following Hurricane Helene, disaggregated by average cost per person (left) and total cost per Zip Code (right). Higher average deprivation costs are located particularly in the northwestern and north-central regions of the state, where average deprivation costs exceed $1,000. These areas may have experienced longer restoration times, greater infrastructure vulnerability, or socioeconomic conditions that limited residents' ability to cope with power outages. Meanwhile, the total deprivation cost map emphasizes the cumulative economic burden across all residents within each Zip code.

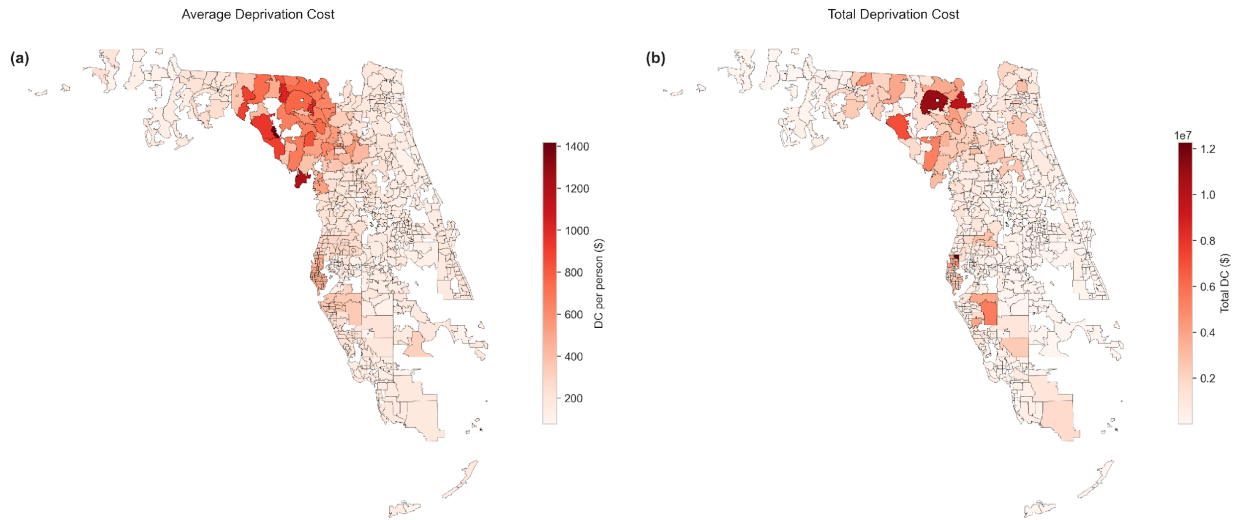

**Figure 3. Spatial distribution of power outage deprivation costs following Hurricane Helene in Florida**. Geographic variation in deprivation costs across Florida Zip codes during Hurricane Helene. (a) Average deprivation cost per person, revealing localized pockets of elevated burden (>$1,000) in northwestern and north-central regions. (b) Total deprivation cost per Zip Code, emphasizing cumulative economic losses with substantial impacts in the northern part.

Figure 4 presents the spatial distribution of deprivation costs across Florida following Hurricane Milton. Compared to Helene, Milton's impact exhibits a broader geographic footprint, with elevated average costs dispersed throughout the central and southwestern regions. Several areas, particularly along the Gulf Coast and in central inland areas, experienced average costs exceeding $700 per person. On the right panel, the total deprivation cost map identifies regional centers where the aggregate economic burden was highest. Notably, a few densely populated counties in central Florida stand out, with total losses surpassing $50 million in some areas. While these locations do not always correspond to the highest per-capita costs, their population scale translates into significant societal impact. The divergence between the two maps highlights the need to interpret deprivation costs through multiple lenses. High per-person costs may signal equity concerns and the need for targeted aid, while large total costs underscore where widespread economic disruptions occurred.

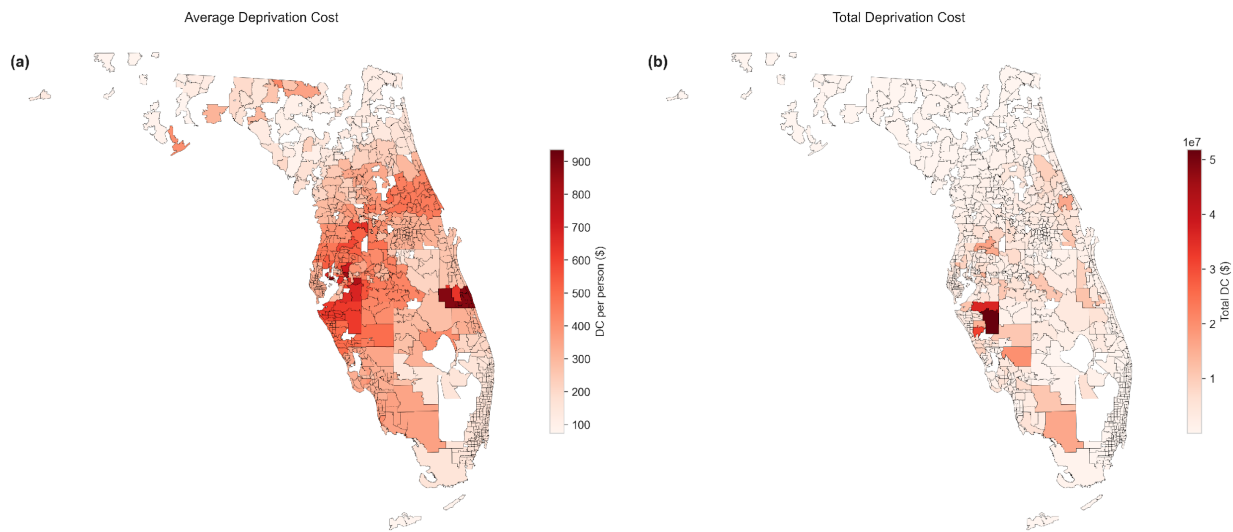

**Figure 4. Spatial distribution of power outage deprivation costs following Hurricane Milton in Florida**. Deprivation cost patterns across Florida following Hurricane Milton, showing a broader geographic impact compared to Hurricane Helene. (a) Average deprivation cost per person, with elevated costs (>$700) dispersed throughout central and southwestern regions along the Gulf Coast. (b) Total deprivation cost per Zip Code, identifying regional centers of highest aggregate burden (>$50 million) in the Tampa Bay area and central Florida, demonstrating the role of population scale in amplifying societal impact.

Figure 5 illustrates the geographic variation in deprivation cost following Hurricane Ida across Louisiana. In contrast to other events, Hurricane Ida generated exceptionally high per-capita deprivation costs, with several southern coastal communities exceeding $5,000 per person. In terms of total deprivation cost, the right panel reveals that the largest aggregate losses are concentrated in more densely populated suburban areas northwest of New Orleans, where costs exceeded $70 million in some areas. This divergence between high average versus high total costs once again underscores the multifaceted nature of disaster impact: areas with concentrated populations may contribute more to overall economic burden, while sparsely populated but highly impacted regions suffer disproportionately at the individual level.

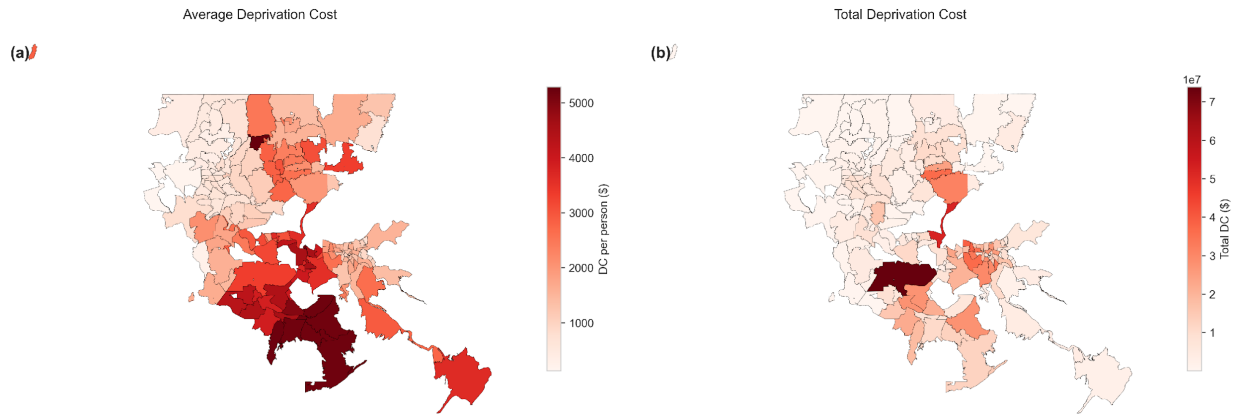

**Figure 5. Spatial distribution of power outage deprivation costs following Hurricane Ida in Louisiana**. Geographic distribution of deprivation costs in Louisiana during Hurricane Ida, showing the most severe individual and community burdens among all studied events. (a) Average deprivation cost per person, with exceptionally high values (>$5,000) in southern coastal communities, indicating prolonged outages and severe vulnerability. (b) Total deprivation cost per Zip Code, revealing peak aggregate losses (>$70 million) in densely populated suburban areas.

Figure 6 through Figure 8 show comprehensive analysis of the relation of deprivation cost to median income, relative restoration rate, and restoration duration. Figure 6 presents a comparative analysis of deprivation cost as a share of household income across four extreme weather events—Beryl, Helene, Milton, and Ida. The y-axis (logarithmic scale) captures the ratio of deprivation cost to income, plotted against median household income on the x-axis. Each data point represents a local unit (i.e., Zip Codes), and the fitted trend lines illustrate different functional relationships for each event. Across all events, a clear inverse relationship is observed: lower-income communities consistently bear a higher burden relative to income, as indicated by elevated deprivation cost shares. For Hurricane Ida, a near-linear decline suggests a persistent disparity across the income spectrum, while events such as Helene and Milton exhibit more pronounced curvature, captured through power-law or log-linear fits. These nonlinear trends reflect steep drops in deprivation burden with modest increases in income, suggesting heightened vulnerability among the lowest-income brackets. Beryl's flatter power fit further reinforces the relatively uniform impact across income levels in that event, though still showing inequality. This visualization reinforces a central finding of the study: deprivation cost is disproportionately affecting households with limited financial resilience. Notably, Ida's data cluster is located at the highest overall values on the vertical axis, indicating substantially higher deprivation cost even for relatively higher-income areas, likely due to the storm's severity and extended outage durations. In contrast, Helene and Beryl were both best modeled with a power-law function, displaying strong curvature that reflects a steeper decline in cost burden at the lower end of the income distribution. This pattern suggests that small increases in income among lower-income households are associated with substantial reductions in relative deprivation, highlighting their heightened vulnerability. Milton's data, best fit by a log-linear function, exhibits a moderate gradient—indicating that while deprivation cost declines with income, the rate of change is less pronounced than in the other events. The vertical separation between event-specific trendlines also reveals meaningful inter-event differences. For a given income level, deprivation cost shares under Ida are roughly an order of magnitude greater than those under Helene, while Beryl and Milton occupy intermediate positions. These differences likely reflect a combination of event intensity, outage duration, local preparedness, and demographic exposure.

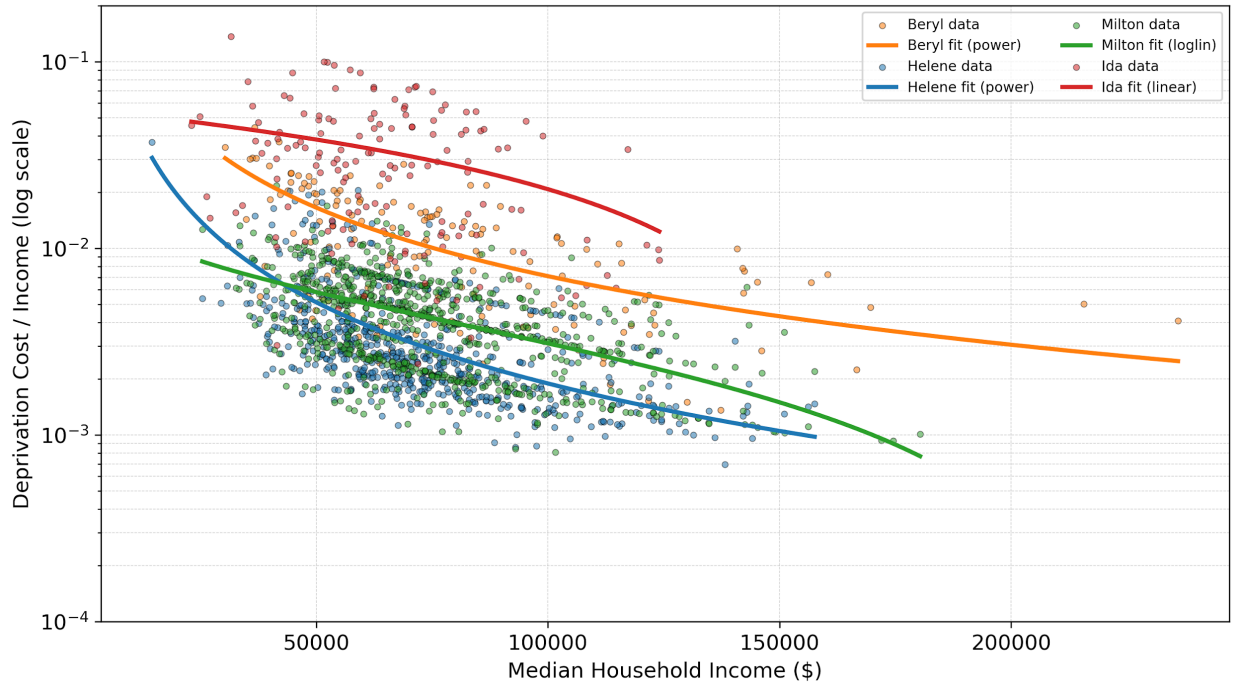

**Figure 6. Cross-event comparison of deprivation cost share by household income**. The logarithmic y-axis shows the deprivation cost ratio plotted against median household income, with fitted curves representing different functional relationships for each event. All events demonstrate inverse relationships with steeper declines for lower-income households, indicating regressive burden distribution. Hurricane Ida shows the highest overall burden levels, while curve shapes reveal event-specific vulnerability patterns.

Similarly, Figure 7 examines the relationship between relative restoration rate and deprivation cost. Across all four hurricane events, a consistent negative association is observed between relative restoration rate and deprivation cost—communities that experienced slower-than-average power restoration incurred significantly higher monetary losses. Ida, fit by a power-law model, exhibits the steepest decline in deprivation cost as restoration rate improves, with localized costs exceeding $10,000 in areas with the slowest recovery. This indicates high sensitivity of deprivation cost to restoration disparities during this severe event. Beryl also shows a strong exponential decay pattern, suggesting that even modest improvements in restoration rate could have yielded substantial reductions in societal costs. Milton and Helene exhibit comparatively flatter curves, with Milton best described by a logarithmic-linear relationship and Helene by a power-law fit. These patterns suggest that while faster restoration is still beneficial, the marginal gains in reducing deprivation are smaller than those observed for Ida or Beryl. Notably, Milton's curve extends further along the x-axis, indicating a broader range of restoration experiences among affected communities.

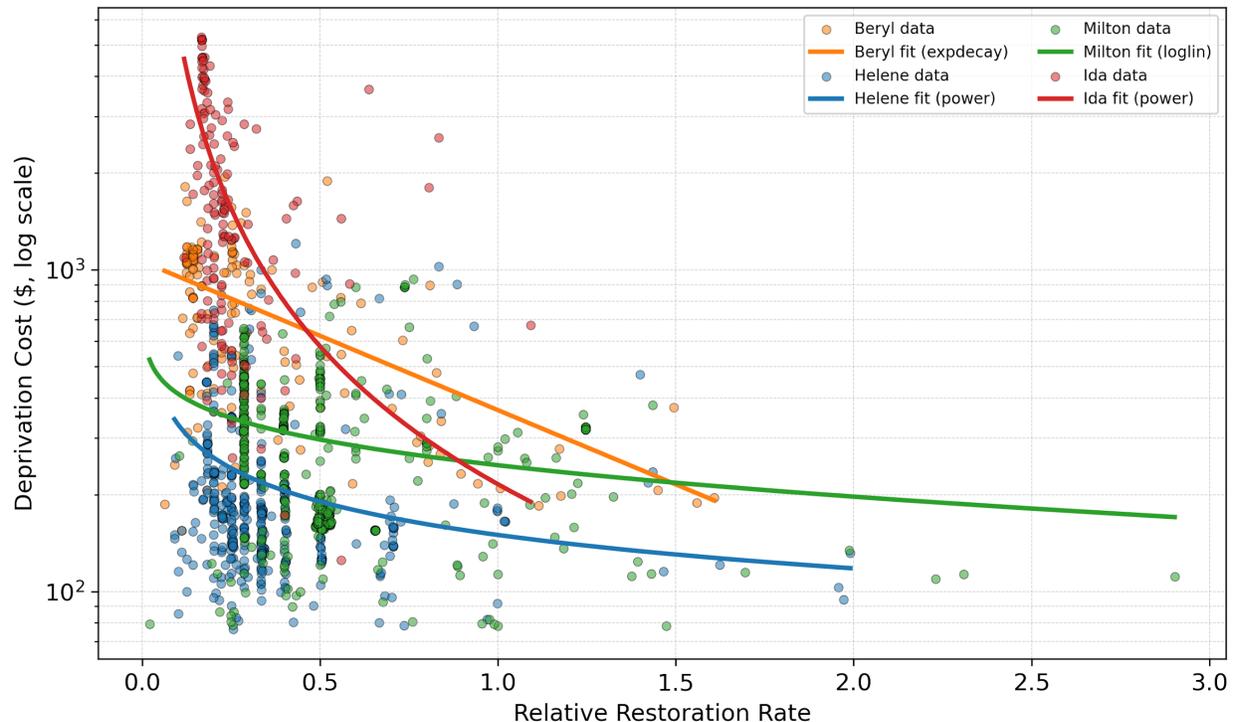

**Figure 7. Relationship between relative restoration rate and deprivation cost across hurricane events**. Association between restoration performance and societal burden across four hurricane events, with deprivation cost plotted on logarithmic scale against relative restoration rate. Fitted curves selected via Akaike Information Criterion show consistent negative relationships, indicating that slower restoration rates correspond to higher deprivation costs. Hurricane Ida exhibits the steepest decline (power-law), while Milton and Helene show more moderate relationships, suggesting varying sensitivity to restoration delays across events.

To further investigate how restoration delays influence societal losses, we analyzed deprivation cost estimates as a function of restoration duration across the four hurricane events: Beryl, Helene, Milton, and Ida (Figure 8). The results indicate that restoration duration is positively associated with deprivation cost, but the strength and form of this relationship vary significantly across events. Hurricane Ida shows a strong positive linear trend, suggesting that longer restoration periods directly translate into higher societal losses. This pattern may reflect prolonged base outages, greater community vulnerability, or delayed recovery operations in high-impact areas. In contrast, Beryl and Helene exhibit flatter exponential relationships, indicating that deprivation costs rose more gradually with increasing restoration durations, possibly due to more uniform or less severe disruptions. Milton shows a mild power-law curve, reflecting a moderate increase in deprivation cost as restore duration lengthens. These findings highlight that longer restoration periods are generally associated with greater deprivation costs, but the relationship is not uniform across events. In some cases, such as Hurricane Ida, the cost burden grows sharply with delay, underscoring the disproportionate impact on already vulnerable communities.

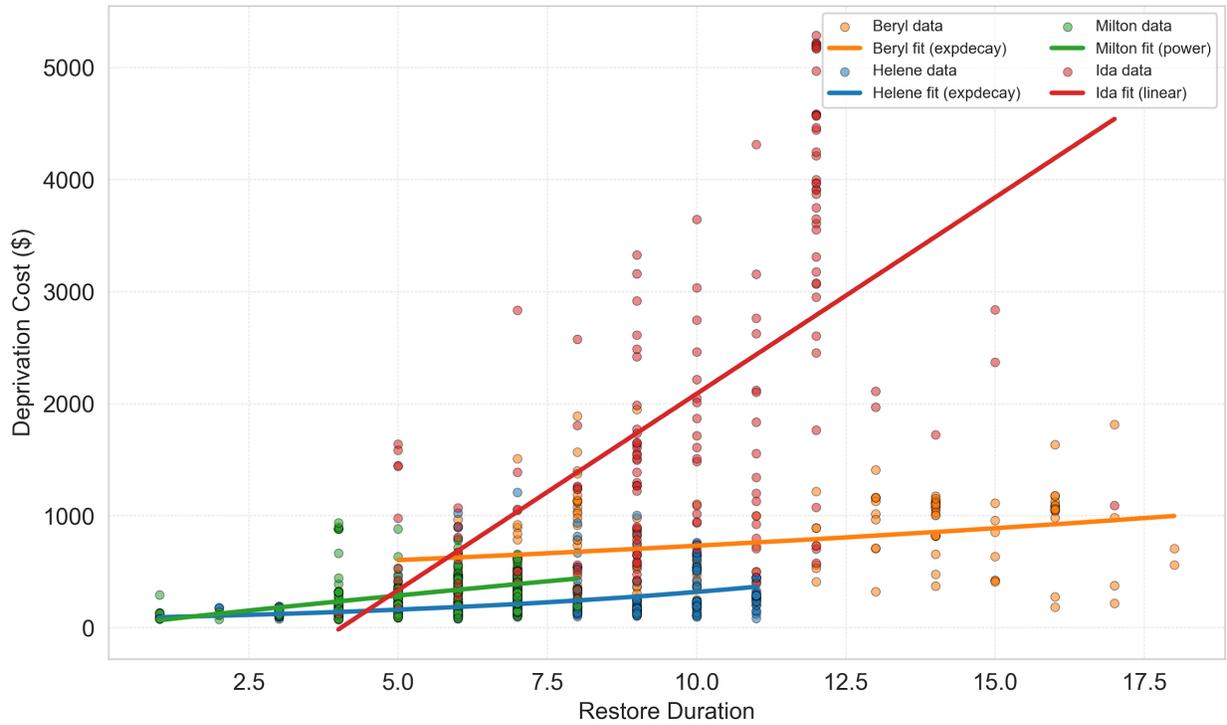

**Figure 8. Deprivation cost as a function of restoration duration across hurricane events**. Relationship between restoration duration and deprivation cost across four hurricane events, with restoration duration defined as days between restoration onset and full recovery. Fitted models selected via Akaike Information reveal positive associations with varying functional forms: Hurricane Ida shows strong linear growth, while Beryl and Helene exhibit flatter exponential relationships. The varying slopes indicate different sensitivities to restoration delays across events and regions.

To elucidate the heterogeneous effects of key predictors on deprivation cost across socio-economic strata, we conducted a SHAP-based Random Forest analysis. The results reveal distinct patterns of feature importance and directional influence between low-income and high-income populations. In both groups, restoration duration emerges as the dominant driver of deprivation cost; however, its marginal impact is substantially amplified in higher-income communities, suggesting heightened sensitivity to prolonged outages. Notably, in the high-income group, the relative restoration rate exhibits a pronounced negative contribution, indicating that delays in comparative restoration performance significantly exacerbate perceived loss. In contrast, the same variable in the low-income cohort shows only modest influence. Median income itself has limited predictive power within each subgroup, underscoring the relative importance of service restoration dynamics over static socio-economic status in shaping deprivation outcomes.

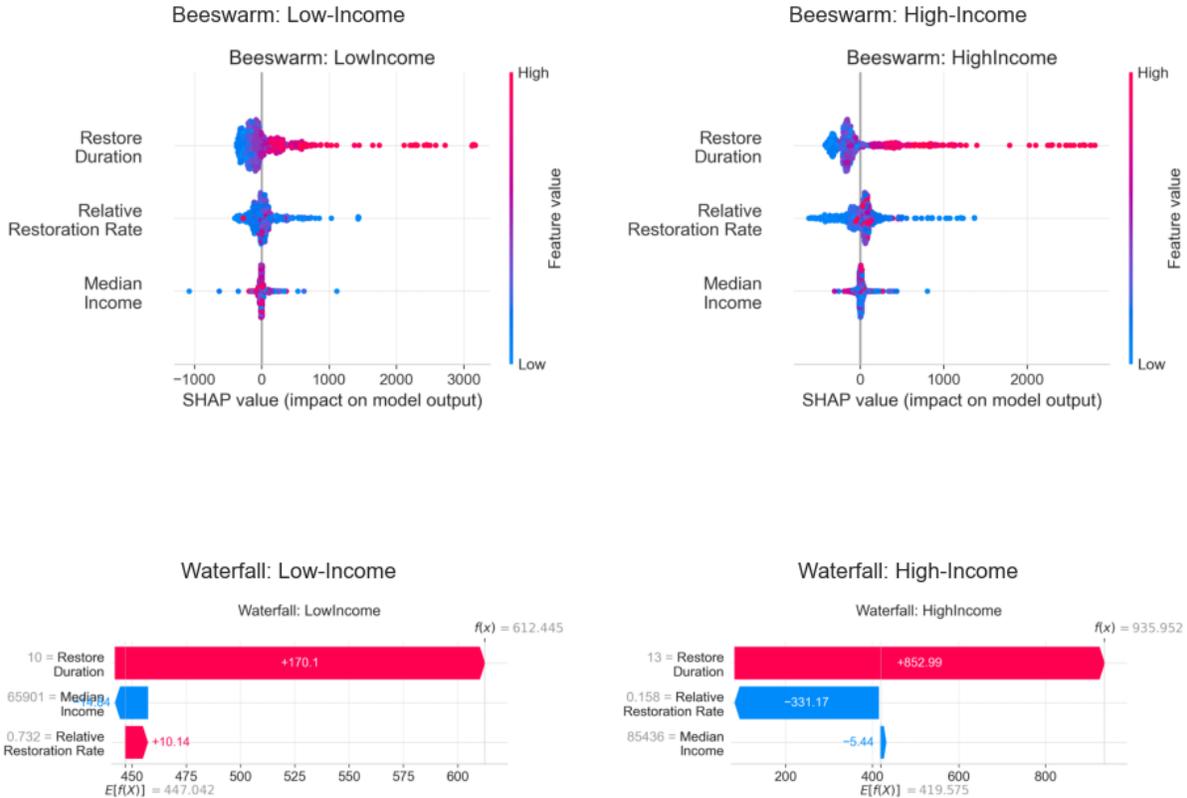

Figure 9. **SHAP analysis of feature importance of deprivation cost in different income**. Results show differential feature influence between low-income and high-income populations, with restoration duration emerging as the dominant predictor in both groups but showing amplified impact in higher-income communities.

Before clustering, to ensure the suitability of input features for unsupervised clustering, we examined the pairwise Pearson correlation coefficients among three key variables—relative restoration rate, restoration duration, and deprivation cost—across four hurricane events (Beryl, Helene, Milton, and Ida). As shown in Figure 10, while some moderate correlations are observed (e.g., between restoration duration and deprivation cost in Ida: r = 0.57; and in Helene: r = 0.43), the overall magnitudes remain well below multicollinearity thresholds. Notably, the relative restoration rate exhibits weak or moderate inverse correlations with deprivation cost (ranging from -0.13 to -0.48), indicating that faster recovery tends to be associated with lower societal burden, though the strength of this relationship varies by storm. Importantly, no pair of variables demonstrates excessively high correlation[37], supporting the assumption that each metric captures distinct information about the recovery process. This statistical independence justifies their joint inclusion in subsequent K-means clustering to identify regional recovery typologies.

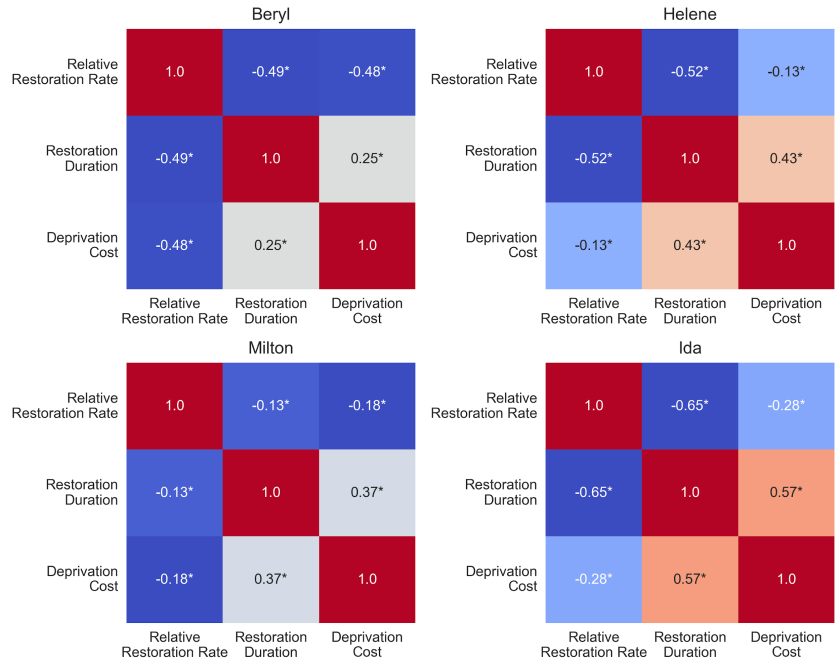

**Figure 10. Correlation matrix of clustering variables across hurricane events**. Pearson correlation coefficients among key variables used in K-means clustering analysis across four hurricane events. The heatmaps show moderate correlations (|r| < 0.7) between restoration duration, relative restoration rate, and deprivation cost variables, confirming low multicollinearity and suitability for clustering analysis. The varying correlation patterns across events reflect different outage and recovery dynamics.

Figure 11 displays the spatial distribution of the resulting four clusters across Harris County and the box plot of the features. Each cluster represents a distinct typology of socio-technical experience during power restoration: Cluster 0 (Red) – High deprivation, slow recovery: These communities faced the highest deprivation costs, the longest outage durations, and the lowest relative restoration rates, while also exhibiting low to middle income levels, representing areas most severely affected by prolonged outages and limited system responsiveness. Cluster 1 (Blue) – Moderate deprivation, highest income: This group experienced moderate deprivation costs, shorter outage durations than Cluster 0, and somewhat improved restoration rates, coinciding with the highest income neighborhoods and reflecting stronger buffering capacity but not the fastest recovery. Cluster 2 (Orange) – High deprivation despite moderate duration: These areas incurred elevated deprivation costs even with moderate outage durations and moderate income levels, suggesting heightened vulnerability per unit of disruption. Cluster 3 (Green) – Low deprivation cost, fast recovery: This cluster experienced the lowest deprivation costs, shortest outage durations, and fastest restoration rates, yet it corresponded to the lowest income neighborhoods, revealing a counterintuitive pattern in which rapid recovery did not align with socioeconomic advantage.

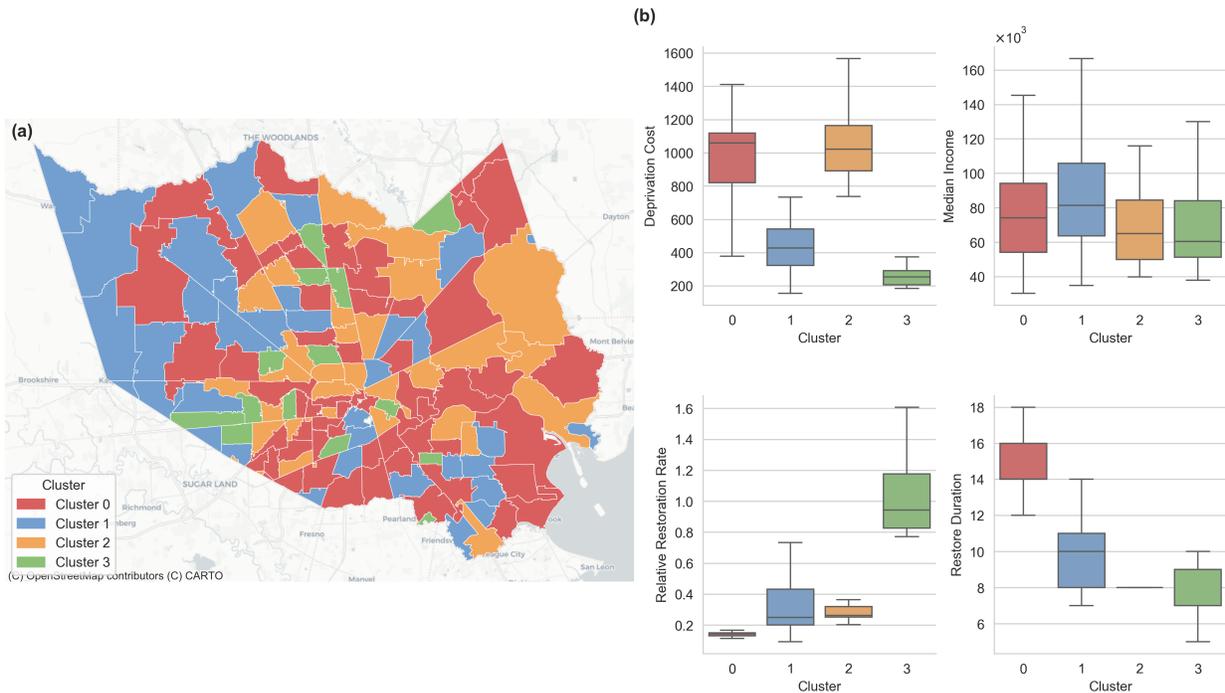

**Figure 11. K-means clustering results for Hurricane Beryl in Harris County**. (a) Geographic distribution showing cluster assignments across Zip Codes. (b) Box plots of standardized features revealing distinct cluster profiles: Cluster 0 (red) represents high deprivation with slow recovery; Cluster 1 (blue) shows moderate burden with moderate income; Cluster 2 (orange) exhibits high deprivation despite low duration; Cluster 3 (green) indicates high income with fast recovery.

Figure 12 shows the case of Helene in Florida that clustered into four distinct groups based on deprivation cost, median income, restoration duration, and relative restoration rate. Cluster 0 (blue) is widely distributed across central and eastern Florida and is characterized by relatively short restoration durations (about 6 days) and moderate deprivation costs (about $150), despite moderate income levels. Cluster 1 (red), concentrated in northern inland areas, represents the most vulnerable group, with the highest deprivation costs and longest restoration durations, while also exhibiting relatively low-income levels. This group likely reflects communities facing structural and infrastructural disadvantages. Cluster 2 (orange) displays moderate deprivation costs and restoration times, forming a transitional group with slightly higher income than Cluster 1 but lower than Cluster 0. In contrast, Cluster 3 (green) encompasses high-performing areas mainly in the southern Florida. These areas feature the highest relative restoration rates, shortest restore durations, and the lowest deprivation costs, suggesting both infrastructure robustness and effective restoration planning.

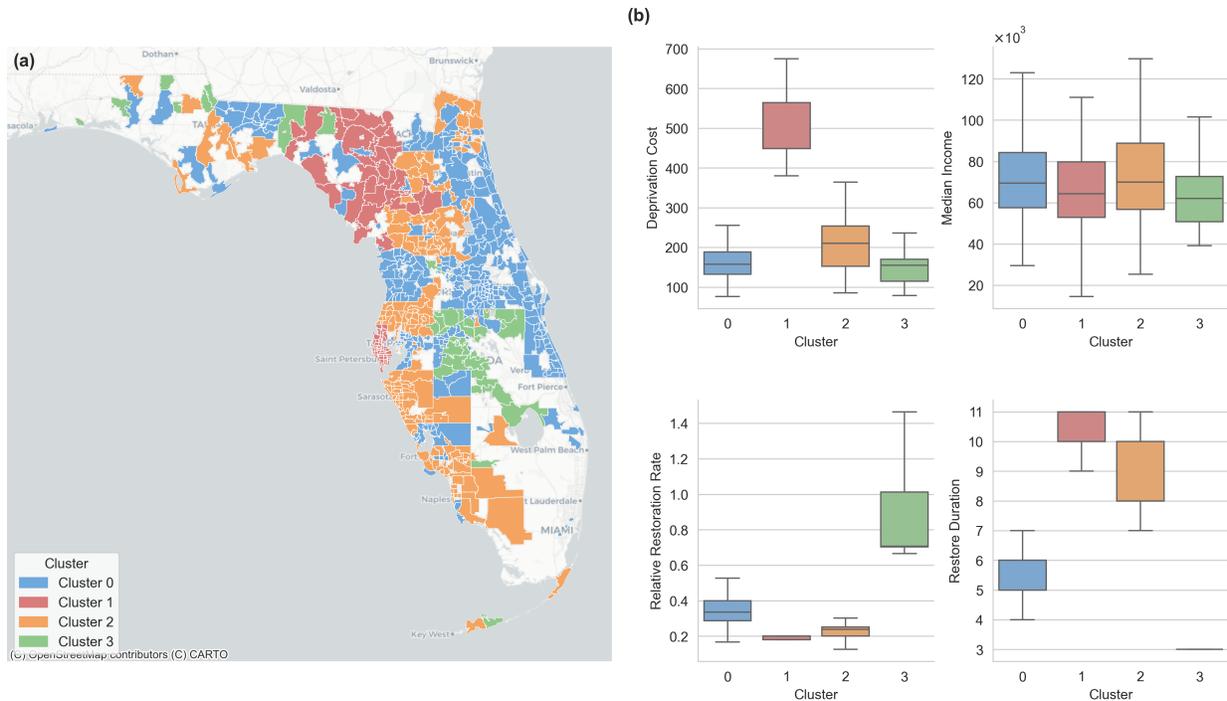

Figure 12. **K-means clustering results for Hurricane Helene in Florida**. (a) Spatial distribution with Cluster 1 (red) was concentrated in vulnerable northern inland areas, Cluster 3 (green) along high-performing Gulf Coast regions, and Clusters 0 and 2 distributed across central and southern areas. (b) Box plots confirming distinct feature distributions with clear separation across deprivation cost, income, and restoration metrics.

The spatial distribution of clusters in Figure 13 reveals strong geographic variation in power restoration outcomes across Florida following Hurricane Milton. Cluster 1 (red) is the most widespread, covering much of the interior and western parts of the state, including areas around Tallahassee and central Florida. These areas experienced the most adverse conditions, with high deprivation costs, long restoration durations, and low relative restoration rates, combined with below-average median income levels. Cluster 2 had the shortest restoration durations (median around 2 days), the highest relative restoration rates (greater than 2.0), and the lowest deprivation costs (around $100). These areas also had relatively high-income levels, suggesting stronger infrastructure and faster system response. Cluster 0 (blue) appears across northern and southern Florida, particularly around Miami-Dade. This cluster had moderate deprivation costs (around $180), shorter outage durations (around 4 days), and low restoration rates. The median income was higher than Clusters 1 and 3, indicating some economic resilience despite slower recovery. Cluster 3 (orange) is scattered across parts of central and northern Florida. It reflects transitional performance, with moderate deprivation costs (around $300), average restore durations (around 5 days), and restoration rates slightly above 1.0. Median income levels were slightly below average.

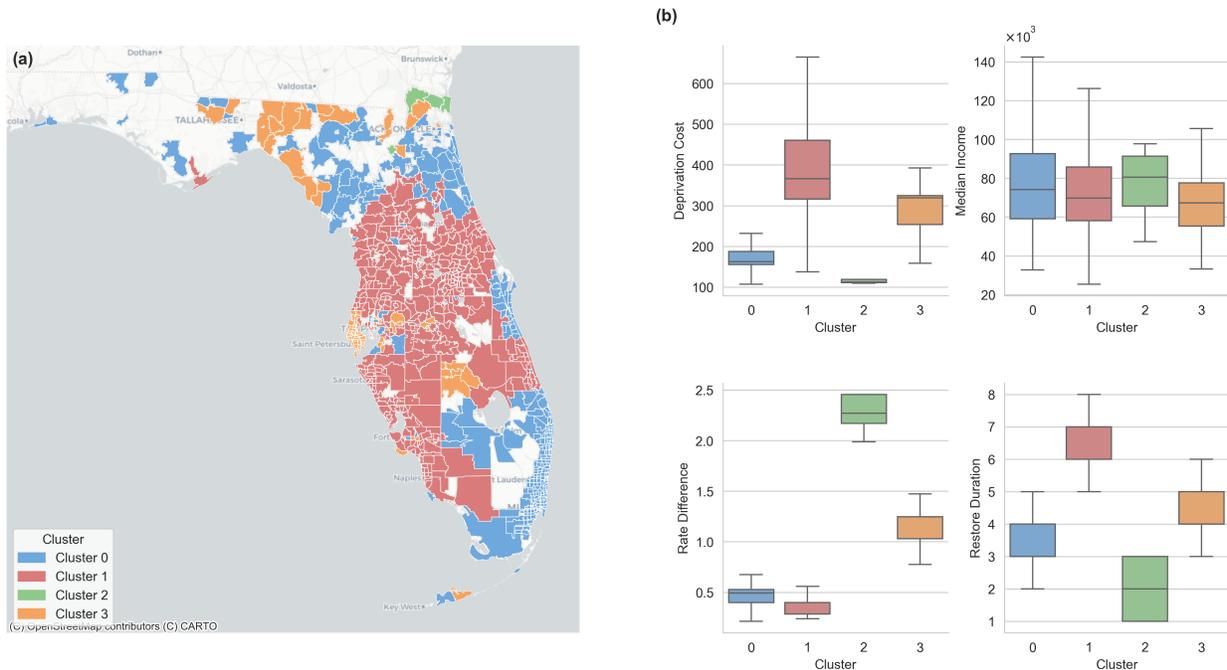

Figure 13. **K-means clustering results for Hurricane Milton in Florida**. (a) Spatial distribution showing widespread Cluster 1 (red) coverage in interior regions with adverse conditions, concentrated Cluster 2 (green) in high-performing areas, and a mixed distribution of moderate-performing clusters. (b) Box plots demonstrating clear cluster differentiation across all features, with Cluster 1 showing the highest deprivation and Cluster 2 showing optimal performance.

Figure 14 presents the spatial distribution of four clusters derived from Hurricane Ida's impact in southern Louisiana, alongside the corresponding boxplots of deprivation cost, median income, restoration duration, and relative restoration rate. Cluster 1 (red) is spatially concentrated across southern coastal parishes. It represents the most disadvantaged group, with the highest deprivation costs, longest restoration durations, and the lowest restoration rates, although with high income level. Cluster 0 (blue) covers broad areas of southeastern and central parishes. It exhibits moderate deprivation costs, long duration, and low restoration rates, suggesting that even relatively affluent communities were not fully insulated from prolonged disruptions. Median income is the lowest, reflecting vulnerable socioeconomic conditions that may have exacerbated sensitivity to outages. Cluster 2 (Orange) represents a mid-performing group with moderate deprivation costs, shorter restoration durations, and higher restoration rates. These areas demonstrate relatively better resilience, likely due to more effective restoration logistics or prioritization in service recovery, and they also correspond to the second-highest income levels, which may have provided additional buffering capacity. Cluster 3 (green), found in isolated pockets, reflects the most efficient recovery. These areas had the shortest restoration durations, lowest deprivation costs, and relatively higher restoration rates, despite modest income levels. This combination suggests favorable infrastructure or prioritization that reduced outage impact even in lower-resource areas. The boxplots in Figure 14 confirm strong separation between clusters across all four dimensions.

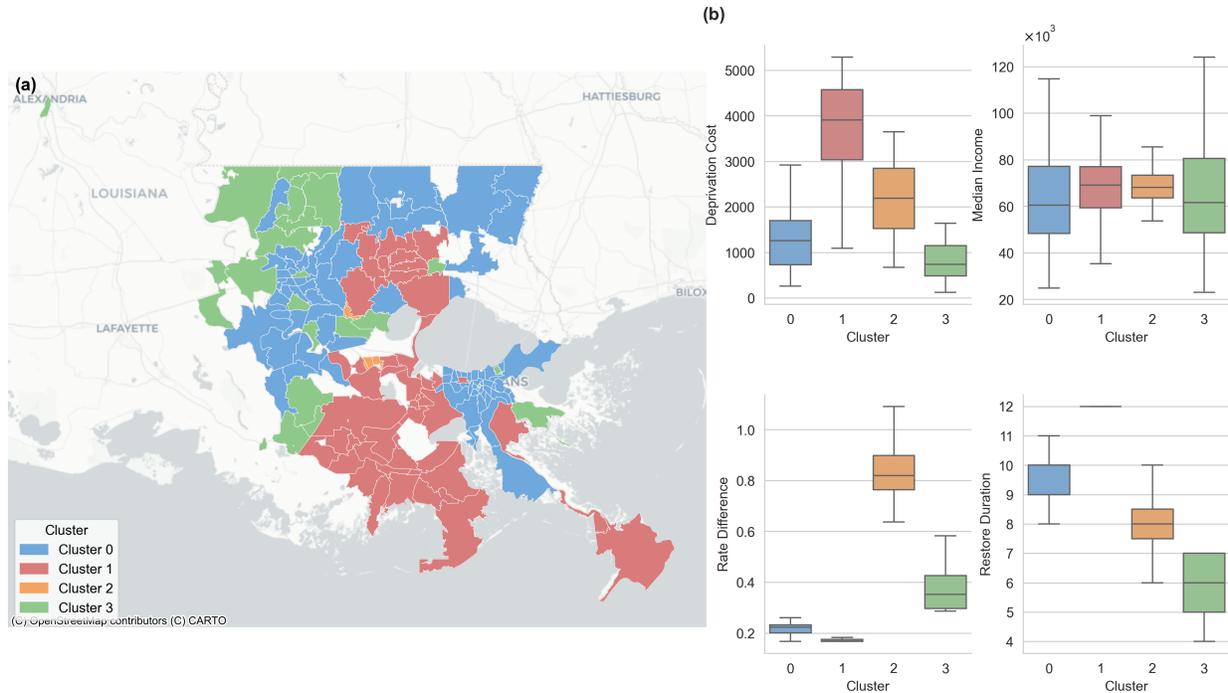

**Figure 14. K-means clustering results for Hurricane Ida in Louisiana** (a) Geographic distribution showing Cluster 1 (red) concentration in vulnerable coastal parishes, Cluster 0 (blue) across southeastern regions, and scattered distribution of moderate and high-performing clusters. (b) Box plots revealing extreme cluster separation with Cluster 1 exhibiting exceptionally high deprivation costs (>$4,000) and longest restoration times, while Cluster 3 (green) shows optimal recovery performance despite modest income levels.

**Discussion**

This study introduces a novel framework for quantifying societal costs of power outages by integrating customer-weighted duration metrics with spatially explicit deprivation cost analysis across four major hurricane events. Our methodology advances beyond traditional utility performance metrics by translating outage exposure into monetary losses through empirically derived willingness-to-pay functions, enabling direct comparison of equity outcomes across different regions and events. The key innovation lies in combining high-resolution temporal outage data with ZIP code-level demographic information to reveal systematic patterns of vulnerability that conventional aggregate metrics completely overlook. By processing sequential outage observations to calculate cumulative customer exposure and converting this to deprivation costs, we provide the first systematic comparison of outage equity outcomes across different storm characteristics and regional contexts.

Our analysis of four hurricanes reveals substantial heterogeneity in societal outage costs, both across events and within states. Hurricane Ida in Louisiana imposed the highest burden, with total deprivation costs reaching $1.50 billion and per-capita costs of $1,757. Milton in Florida followed with $1.26 billion total ($387 per capita), while Beryl in Harris County, Texas resulted in $629 million ($674 per capita), and Helene in Florida generated $411 million ($285 per capita). These spatial patterns highlight a critical distinction between equity and aggregate impact metrics. Areas experiencing the highest per-capita deprivation costs—signaling equity concerns—often differ from those driving the largest total costs, which represent aggregate social losses. This divergence reveals how different analytical lenses capture distinct dimensions of hurricane impacts. A consistent regressive pattern emerges across all events: deprivation cost shares decline with income. However, the functional form of this relationship varies by storm, from near-linear for Ida to power-law or log-linear for others. Notably, Ida consistently imposed the highest burden levels at any given income level, demonstrating how both storm severity and community capacity determine the distribution of outage costs. Restoration dynamics emerge as the primary mechanistic driver of these costs. Longer restoration durations and lower relative restoration rates strongly correlate with higher deprivation costs. Explainability analysis confirms restoration duration as the dominant predictor, while faster restoration rates provide protective effects, particularly in higher-income areas. Notably, income itself shows limited predictive power within income strata once restoration metrics are accounted for. Our unsupervised clustering analysis identifies four distinct recovery typologies, ranging from high-deprivation, slow-recovery zones (predominantly lower-income) to high-income, fast-recovery areas. This classification reveals that communities with similar aggregate outage statistics can experience vastly different welfare losses. These findings offer actionable insights for utilities and regulators. Concerning feature contributions to deprivation costs, SHAP analysis reveals restoration duration as the dominant predictor across income strata, with amplified impacts in higher-income communities where prolonged outages generate disproportionately higher perceived losses. Relative restoration rate exhibits pronounced negative contributions primarily affecting affluent areas, indicating that delays in comparative restoration performance significantly exacerbate deprivation in higher-income communities. Power outage characteristics consistently outweigh static socio-demographic features in determining deprivation outcomes, though baseline vulnerability remains elevated in low-income areas. The differential sensitivity patterns suggest that while restoration delays impose costs across all communities, the marginal impact varies systematically with socioeconomic status, creating complex equity implications for restoration prioritization strategies.

Regarding hidden heterogeneity revealed by clustering, spatial analysis identifies distinct community typologies that expose systematic inequities invisible to conventional metrics. These include high-income fast-recovery areas characterized by shortest outage durations, lowest deprivation costs, and highest restoration rates; moderate-burden communities with better economic buffering capacity despite longer restoration times; high-vulnerability zones experiencing elevated deprivation costs despite shorter durations due to limited adaptive capacity; and persistently disadvantaged areas with prolonged outages, elevated costs, and slow restoration rates. These clusters demonstrate that communities with similar aggregate restoration statistics experience vastly different deprivation costs based on underlying socioeconomic conditions and infrastructure characteristics, revealing spatial inequalities that traditional utility performance metrics completely fail to capture. The clustering results challenge conventional assumptions about uniform recovery experiences within affected regions and demonstrate that outage impacts are systematically structured by pre-existing socioeconomic conditions.

**Concluding Remarks**
Unlike system reliability indices that tally interruptions at the system average, this study presents an impact-based evaluation by pricing what people actually lose when the lights go out. By turning sequential outage observations into customer-weighted outage-days, restore duration, and relative restoration rate, and then mapping that exposure to deprivation cost (DC) in dollars, the analysis captures both how long and how costly outages are for different communities—revealing regressive burdens, spatial heterogeneity, and recovery typologies that SAIDI/CAIDI cannot see. The result is a transparent, additive welfare metric that is directly comparable across events (Beryl, Helene, Milton, Ida) and directly usable for prioritization, not just performance reporting. In short, it shifts storm evaluation from *how many customers were out and for how long?* to *who paid how much in social loss—and why*? supplying decision-ready evidence for equity-informed restoration. Because DC is denominated in dollars, it closes the loop to investment decisions: historical storm DC surfaces and the empirically observed links—DC rising with restore duration and falling with relative restoration rate—enable straightforward expected-avoided-loss calculations to justify resilience spending. Practically, utilities can (i) estimate baseline expected DC for plausible future storms using past event relationships, (ii) model how candidate measures that reduce restore duration or increase relative restoration rate would shift those curves in affected ZCTAs or clusters, and (iii) compute avoided social losses to support benefit–cost analysis, performance-based regulation, and equity targeting. This turns restoration speed into a monetized lever ("value of a day saved") and channels capital toward places where each minute restored prevents the greatest social cost, aligning resilience plans with measurable public welfare gains.

Accordingly, the study and its findings advance the field by developing and operationalizing a deprivation cost framework that quantifies the societal impacts of power outages at granular spatial resolution. Our methodology integrates high-resolution temporal outage data with localized socioeconomic information, converting physical service interruptions into economic welfare losses based on household deprivation costs. This approach enables the first systematic cross-event comparison of equity outcomes, demonstrating empirically how outage impacts follow regressive patterns across diverse geographic and storm contexts. Through explainable AI (SHAP analysis) and unsupervised learning (K-means clustering), we reveal the complex relationships between infrastructure performance, restoration dynamics, and social vulnerability—uncovering systemic disparities that conventional approaches miss. From a practical perspective, the framework offers transformative applications for utilities, regulators, and emergency managers. By monetizing the societal burden of outages, utilities can evolve from purely technical restoration strategies to social impact-aware prioritization that minimizes social harm. Regulators gain an empirical foundation for reforming oversight mechanisms, complementing traditional reliability indices with metrics that incentivize impact-focused investments. These findings enable infrastructure modernization decisions that enhance both grid resilience and social impact, directly addressing the disproportionate impacts of extreme weather events.

The following methodological considerations shape the interpretation of our findings. The deprivation cost function, though empirically derived, represents a standardized approach that may not fully reflect variations in willingness-to-pay across diverse cultural, health, and social contexts. Our ZIP code-level analysis provides valuable community-scale insights while aggregating across heterogeneous populations, which may smooth some neighborhood-level vulnerability patterns. The study's temporal focus on acute event impacts captures immediate restoration costs, with longer-term consequences, such as cascading business effects, health outcomes, and secondary infrastructure impacts representing areas for future investigation. These considerations suggest opportunities to extend and refine the framework as additional data and methodological approaches become available.

**Acknowledgement**

This research was supported by the National Science Foundation (NSF) CAREER Award [CMMI-1846069]. The authors gratefully acknowledge the National Renewable Energy Laboratory (NREL) for providing access to the I-EAGLE dataset, which was invaluable for this study.

**References**

1. Gjorgiev, B. & Sansavini, G. Identifying and assessing power system vulnerabilities to transmission asset outages via cascading failure analysis. *Reliab. Eng. Syst. Saf.* **217**, 108085 (2022).

2. Ma, J., Li, B., Omitaomu, O. A. & Mostafavi, A. Establishing Nationwide Power System Vulnerability Index across US Counties Using Interpretable Machine Learning. Preprint at https://doi.org/10.48550/arXiv.2410.19754 (2024).

3. Li, B. & Mostafavi, A. Unraveling fundamental properties of power system resilience curves using unsupervised machine learning. *Energy AI* **16**, 100351 (2024).

4. Poudyal, A., Poudel, S. & Dubey, A. Risk-Based Active Distribution System Planning for Resilience Against Extreme Weather Events. *IEEE Trans. Sustain. Energy* **14**, 1178–1192 (2023).

5. Abidin, Z. & Hadi, A. Analisis System Avarage Interruption Frequency Index dan System Average Interruption Duration Index Beroreantasi Pelanggan Pada Gangguan Jaringan Tegangan Rendah dan Menengah. *INOVTEK - Seri Elektro* **2**, 63 (2020).

6. Eto, J. H., LaCommare, K. H., Caswell, H. C. & Till, D. Distribution system versus bulk power system: identifying the source of electric service interruptions in the US. *IET Gener. Transm. Distrib.* **13**, 717–723 (2019).

7. Daeli, A. & Mohagheghi, S. Power Grid Infrastructural Resilience against Extreme Events. *Energies* **16**, 64 (2022).

8. Schumann, Z. D. & Chini, C. M. Component Assessment of the Electric Transmission Grid to Hurricanes. *Earths Future* **11**, e2023EF003525 (2023).

9. Mitsova, D., Esnard, A.-M., Sapat, A. & Lai, B. S. Socioeconomic vulnerability and electric power restoration timelines in Florida: the case of Hurricane Irma. *Nat. Hazards* **94**, 689–709 (2018).

10. Dugan, J., Byles, D. & Mohagheghi, S. Social vulnerability to long-duration power outages. *Int. J. Disaster Risk Reduct.* **85**, 103501 (2023).

11. Shao, J., Wang, X., Liang, C. & Holguín-Veras, J. Research progress on deprivation costs in humanitarian logistics. *Int. J. Disaster Risk Reduct.* **42**, 101343 (2020).

12. Poudineh, R. & Jamasb, T. Electricity Supply Interruptions: Sectoral Interdependencies and the Cost of Energy Not Served for the Scottish Economy. *Energy J.* **38**, 51–76 (2017).

13. Baik, S., Davis, A. L., Park, J. W., Sirinterlikci, S. & Morgan, M. G. Estimating what US residential customers are willing to pay for resilience to large electricity outages of long duration. *Nat. Energy* **5**, 250–258 (2020).

14. Clark, S. S., Peterson, S. K. E., Shelly, M. A. & Jeffers, R. F. Developing an equity-focused metric for quantifying the social burden of infrastructure disruptions. *Sustain. Resilient Infrastruct.* **8**, 356–369 (2023).

15. Klytchnikova, I. & Lokshin, M. Measuring Welfare Gains from Better Quality Infrastructure. *J. Infrastruct. Dev.* **1**, 87–109 (2009).

16. Bhusal, N., Abdelmalak, M., Kamruzzaman, M. & Benidris, M. Power System Resilience: Current Practices, Challenges, and Future Directions. *IEEE Access* **8**, 18064–18086 (2020).

17. Raoufi, H., Vahidinasab, V. & Mehran, K. Power Systems Resilience Metrics: A Comprehensive Review of Challenges and Outlook. *Sustainability* **12**, 9698 (2020).

18. Xie, D., Cai, S. & Gui, X. Reclaiming justice for energy-vulnerable populations: Evidence from the city of los angeles. *Energy Strategy Rev.* **51**, 101317 (2024).

19. Ferrall, I., Callaway, D. & Kammen, D. M. Measuring the reliability of SDG 7: the reasons, timing, and fairness of outage distribution for household electricity access solutions. *Environ. Res. Commun.* **4**, 055001 (2022).

20. Dargin, J. S. & Mostafavi, A. Human-centric infrastructure resilience: Uncovering well-being risk disparity due to infrastructure disruptions in disasters. *PLOS ONE* **15**, e0234381 (2020).

21. Li, X., Ma, J. & Mostafavi, A. Anatomy of a Historic Blackout: Decoding Spatiotemporal Dynamics of Power Outages and Disparities During Hurricane Beryl. Preprint at https://doi.org/10.48550/ARXIV.2501.10835 (2025).

22. Duan, T., Zhang, F. & Li, Q. Uncovering cooling center usage as an adaptation strategy for hurricane-blackout-heat compound hazards during Hurricane Beryl (2024). *Npj Nat. Hazards* **2**, 53 (2025).

23. Amorim, R., Villarini, G., Czajkowski, J. & Smith, J. Flooding from Hurricane Helene and associated impacts: A historical perspective. *J. Hydrol. X* **27**, 100204 (2025).

24. Hagen, A. B. The 2024 Atlantic Hurricane Season: Helene and Milton Highlight Five U.S. Hurricane Landfalls. *Weatherwise* **78**, 19–28 (2025).

25. LeComte, D. U.S. Weather Highlights: 2021 Hurricane Ida, Historic Cold, Heat, Drought, and Severe Storms. *Weatherwise* **75**, 14–23 (2022).

26. U.S. Census Bureau. 2018-2022 American Community Survey 5-Year Estimates. (2022).

27. U.S. Census Bureau. TIGER/Line Shapefiles. (2024).

28. Cantillo, V., Serrano, I., Macea, L. F. & Holguín-Veras, J. Discrete choice approach for assessing deprivation cost in humanitarian relief operations. *Socioecon. Plann. Sci.* **63**, 33–46 (2018).

29. Holguín-Veras, J. *et al.* Econometric estimation of deprivation cost functions: A contingent valuation experiment. *J. Oper. Manag.* **45**, 44–56 (2016).

30. Macea, L. F., Amaya, J., Cantillo, V. & Holguín-Veras, J. Evaluating economic impacts of water deprivation in humanitarian relief distribution using stated choice experiments. *Int. J. Disaster Risk Reduct.* **28**, 427–438 (2018).

31. Wang, X., Wang, X., Liang, L., Yue, X. & Van Wassenhove, L. N. Estimation of Deprivation Level Functions using a Numerical Rating Scale. *Prod. Oper. Manag.* **26**, 2137–2150 (2017).

32. Li, X. *et al.* Estimating Deprivation Cost Functions for Power Outages During Disasters: A Discrete Choice Modeling Approach. Preprint at https://doi.org/10.48550/ARXIV.2506.16993 (2025).

33. Yamaoka, K., Nakagawa, T. & Uno, T. Application of Akaike's information criterion (AIC) in the evaluation of linear pharmacokinetic equations. *J. Pharmacokinet. Biopharm.* **6**, 165–175 (1978).

34. Likas, A., Vlassis, N. & J. Verbeek, J. The global k-means clustering algorithm. *Pattern Recognit.* **36**, 451–461 (2003).

35. Zhao, J., Lee, C.-D., Chen, G. & Zhang, J. Research on the Prediction Application of Multiple Classification Datasets Based on Random Forest Model. in *2024 IEEE 6th International Conference on Power, Intelligent Computing and Systems (ICPICS)* 156–161 (IEEE, Shenyang, China, 2024). doi:10.1109/ICPICS62053.2024.10795875.

36. Hamilton, R. I. & Papadopoulos, P. N. Using SHAP Values and Machine Learning to Understand Trends in the Transient Stability Limit. Preprint at https://doi.org/10.48550/ARXIV.2302.06274 (2023).

37. Toubiana, D. & Maruenda, H. Guidelines for correlation coefficient threshold settings in metabolite correlation networks exemplified on a potato association panel. *BMC Bioinformatics* **22**, 116 (2021).